\documentclass[aps,prl,twocolumn,preprintnumbers,amsmath,amssymb,floatfix,groupedaddress]{revtex4}

\usepackage{enumerate}
\usepackage{graphicx}
\usepackage{color}
\usepackage[utf8]{inputenc}
\usepackage{enumitem}
\usepackage{appendix}

\usepackage{fontenc}  
\usepackage{amsmath}

\usepackage[francais]{babel}
\newcommand{\beq}{\begin{equation}}
\newcommand{\eeq}{\end{equation}}
\newcommand{\bea}{\begin{eqnarray}}
\newcommand{\eea}{\end{eqnarray}}
\newcommand{\ba}{\begin{array}}
\newcommand{\ea}{\end{array}}

\newcommand{\bef}{\begin{figure}}
\newcommand{\eef}{\end{figure}}

\begin{document}

\title{Contextual inferences, nonlocality, \\ and the incompleteness of quantum mechanics. }

\author{Philippe Grangier}

\affiliation{ \vskip 2mm
Laboratoire Charles Fabry, Institut d'Optique Graduate School, 
Centre National de la Recherche Scientifique, Universit\'e Paris~Saclay, F91127 Palaiseau, France.}

\begin{abstract}
It is known that  ``quantum non locality'', leading to the violation of Bell's inequality and more generally of classical local realism, can be attributed to the conjunction of two properties, that we call here elementary locality and predictive completeness. Taking this point of view, we show again that quantum mechanics violates predictive completeness, allowing to make contextual inferences, which can in turn explain why quantum non locality does not contradict relativistic causality. But if the usual quantum state $\psi$ is predictively incomplete, how to complete it ? We give here a set of new arguments to show that $\psi$ should be completed indeed, not by looking for any ``hidden variables", but rather by specifying the measurement context, which is required to define actual probabilities  over a set of mutually exclusive physical events.

\end{abstract}

\maketitle


\noindent {\it Introduction.--} 
After many years of theoretical and experimental research, it can be now said that the door has been closed on the historical Einstein and Bohr's  quantum debate \cite{Aspect,EPR,Bohr}. On its way, this research opened the door to many new ideas and experiments, leading ultimately to the development of quantum technologies. As a reflection on these evolutions, our point view here is to go back to the Einstein-Bohr debate, and to propose answers to the initial questions: is the ``wave function" a complete description of physical reality ? what is the role of locality ? what about relativistic causality ? We will see that contrary to what is often said, Einstein, Podolsky and Rosen were maybe not so wrong, and Bohr not so right - and that some lesson may be learnt about what Quantum Mechanics is telling us on physical reality.

Our reasoning will use the idea of contextuality, which is currently an extremely active field of research, connected with many foundational issues \cite{Spekkens,Oldo,Abramsky,Cabello,Acin,Dzhafarov,DAriano,Raussendorf}. But rather than pursuing these interesting lines of research, we will step back to discussions from the 1980's \cite{ETJ,JJ,BJ}, which were maybe too quickly dismissed. This is because fully exploiting them amounts to admit  that the usual $| \psi \rangle$ is incomplete, which is a shocking statement rejected by Bohr himself in 1935 \cite{Bohr}. However, many things have happened since then, especially with regards to contextuality and nonlocality.  So in this Letter propose a ``not so shocking'' way to complete $| \psi \rangle$ : very schematically, it tells that a usual state vector is incomplete as long as the complete set of commuting operators admitting this vector as an eigenstate has not been specified. More details will be given below, as well as how to use this idea for our purpose.  

\noindent {\it Probabilistic framework.--} We will use a general framework for conditional probabilities, as presented for instance by E.T. Jaynes in \cite{ETJ}, and also related to the analysis in \cite{JJ,BJ}. We emphasize that these calculations are quite general, and do not imply any commitment to a specific view on probabilities - Bayesian or otherwise.  The equations we will write apply both to usual quantum mechanics and to local hidden variable theories (LHVT), and the main interest of this calculation is to show explicitly where these two descriptions split, and why \cite{options}. 

We will consider the well-known  EPR-Bohm-Bell scheme \cite{Aspect,Laloe}, where polarizations measurements are carried out on entangled photon pairs, described by some quantity  $\lambda$ in  a variable space $\Lambda$.  Alice and Bob carry out measurements defined by respective polarizers' orientations $x$ and $y$, and get binary results $a=\pm 1$ and $b=\pm 1$.

According to usual rules of probabilities, and  without loss of generality  \cite{ETJ},  one can write the following relation between conditional probabilities, by conditioning on $\lambda$ in some a priori unknown hidden variable space $\Lambda$
\beq
P(ab|xy) = \sum_{\lambda \in \Lambda} P(ab|xy\lambda) P(\lambda|xy) \label{eq1}
\eeq

In addition to this purely probabilistic relation, we have to introduce some requirements about the physics we want to describe, and we will do it in the most general way : we assume that usual Quantum Mechanics (QM) and special relativity in the form of Relativistic Causality (RC) are true.
 We note that being true does not necessarily mean being complete \cite{Laloe,Peres,completing}, and we will come back to that issue later on. 
 
 It should be clear also that theories where $a$ and $b$ are deterministic functions of $\lambda$, $x$, $y$ do fit in this probabilistic framework as special cases; however determinism has important consequences, to be discussed below.
\vskip 2 mm

\noindent {\it Enforcing relativistic causality.--}
A first consequence of RC, sometimes called ``freedom of choice'', consists in requiring that $\lambda$ does not depend on the  variables ($x$, $y$) representing Alice and Bob's choices of  measurement settings. In other words,  the choices of measurements ($x$, $y$) should not act on the way photons are emitted ($\lambda$),  since these events are space-like separated. This boils down to the independence condition  $P(\lambda|xy) = P(\lambda)$, or equivalently $P(xy|\lambda) = P(xy)$
which is fulfilled by all the theories we are interested in \cite{superdet}, so that
\beq
P(ab|xy) = \sum_{\lambda \in \Lambda} P(ab|xy\lambda) P(\lambda)  \label{eq2}
\eeq

For a given initial state $\lambda$ of the pair a relevant theory should provide $P(ab|xy\lambda)$, so we will now focus on this conditional probability.  For the sake of clarity, $\lambda$ is a generic notation to specify whatever may be specified about the emission of the photon pair, in a given shot. This may include variables that fluctuate from shot to shot, and other variables that don't. On the other hand, $x$, $y$ and $\lambda$ are not causally related as written above.

Note that eq. (\ref{eq2}) is true also for QM, where the variable space $\Lambda$ contains only one $\lambda$ corresponding to the initial state of the entangled pair (e.g. a singlet state).  It is standard in recent demonstrations of Bell's inequalities \cite{Bell,BCH} to assume that $P(ab|xy\lambda)$ is a probability for a given $\lambda$, so there is no restriction of generality here. 

Now, without any further assumptions, one can always write from basic rules of inference \cite{ft1} 
\bea
P(ab|xy\lambda) &=& P(a|xy\lambda)P(b|xy\lambda a) \nonumber \\
&=& P(a|xy\lambda b) P(b|xy\lambda)  \label{eq3}
\eea
where the two decompositions refer respectively to Alice and Bob, and on Alice's side 

$\bullet \; \;$ $P(a|xy\lambda) =$ probability of Alice getting result $a$ for input $x$

$\bullet \; \;$ $P(b|xy\lambda a) =$ probability of Bob getting result $b$ for input $y$, calculated by Alice who knows $x$ and $a$

\noindent whereas on Bob's side 

$\bullet \; \;$ $P(b|xy\lambda) =$ probability of Bob getting result $b$ for input $y$

$\bullet \; \;$ $P(a|xy\lambda b) =$ probability of Alice getting result $a$ for input $x$, calculated by Bob who knows $y$ and $b$

Clearly a meaningful requirement in eq. (\ref{eq3}), again related to RC,  is that the choice of measurement by Alice (resp. Bob) should not have an influence on the result by Bob (resp. Alice). This implies that 
$P(a|xy\lambda) = P(a|x\lambda)$ and $P(b|xy\lambda) = P(b|y\lambda)$, and we will call this condition ``elementary locality" (EL), meaning that it is fulfilled for each given $\lambda$. 
As a consequence one has 
\bea
P(ab|xy\lambda) &=& P(a|x\lambda) P(b|y\lambda,xa) \nonumber \\
&=& P(a|x\lambda,yb) P(b|y\lambda) \label{eq4}
\eea
where in general one cannot remove $xa$ from $P(b|y\lambda,xa)$, nor $yb$ from $P(a|x\lambda,yb)$. Let us emphasize that so far we have respected QM and RC at each step, and it can easily be checked that  Bell's inequalities cannot  be obtained from eq.(\ref{eq4}).   Correspondingly,  if interpreted ``\`a la Bell'', keeping $xa$ and $yb$ in eq.(\ref{eq4}) looks like an influence of one measurement on the other side. Yet, this conclusion is not warranted since Alice calculates a probability for Bob's result, by using only what is locally available to her (resp. him by switching  Alice and Bob); this does not influence in any way what is happening on the other side. 
\vskip 2 mm

\noindent{\it Contextual inferences vs Bell's hypotheses.--}
We conclude that  given eq.(\ref{eq4}) there is still a missing step to reach Bell's theorem. In order to identify it, let us recall  that locality ``\`a la Bell'' can be seen as a conjunction of two conditions \cite{ft2}.  The first condition is ``elementary locality" (EL), already spelled out  above: 
\vskip 2 mm

(EL)  $P(a|xy\lambda) = P(a|x\lambda)$ and $P(b|xy\lambda) = P(b|y\lambda)$  and  it is verified due to RC as explained before \cite{ft3}.
\vskip 1 mm

\noindent A second condition - let us call it ``predictive completeness" \cite{BJ} is given by : 
\vskip 1 mm

(PC)  $P(a|bxy\lambda) = P(a|xy\lambda)$ and $P(b|axy\lambda) = P(b|xy\lambda)$, and it will be interpreted physically below. 
\vskip 2 mm

Taken together  the conditions (EL)  and (PC) entail the factorization condition $P(ab|xy\lambda) = P(a|x\lambda) P(b|y\lambda)$, and therefore lead to Bell's inequalities \cite{Bell,BCH}. 
\vskip 2 mm

In order to justify the additional hypothesis (PC) and  the wording ``predictive completeness", one must emphasize that Bell's factorization condition $P(ab|xy\lambda) = P(a|x\lambda) P(b|y\lambda)$ relies on the idea that $\lambda$ specifies ``everything which can be known" about the pair of particles; given this assumption, condition (PC) should be obvious, because knowing $xa$ cannot bring anything more to Alice's probability calculation; hence the name of predictive completeness. For instance, theories where $a$ and $b$ are deterministic functions of $\lambda$, $x$, $y$ must satisfy (PC). 
\vskip 1 mm

On the other hand,  $(\lambda x a)$ occurring in the probability $P(b|y\lambda, xa) = P(b|y, \lambda x a)$ is not a property carried by Bob's particle, but it involves both the properties of Bob's particle (included in $\lambda$) and the result of Alice's measurement (described by $xa$). In other words, $(\lambda x a)$ refers to a property of Bob's particle, not in and by itself, but {\it within a context} defined by Alice's result \cite{ft4,CO2002,csm1,csmBell,csm3,csm4a,csm4b,debate,myst}. Alice's context and result cannot have any influence on Bob's particle,  and they don't, since $(\lambda x a)$ is only used locally by Alice, according again to eq.  (\ref{eq4}). 
\vskip 1 mm

Given this situation, we take a major new step beyond previous discussions, by admitting that the description given by $\lambda$ (or $\psi$ in the quantum case) is incomplete indeed, and that knowing $(xa)$ does bring something new to Alice. Then condition (PC) can be violated, by {\bf Alice  making  a  ``contextual inference'' about Bob's result.}
In order to make sense of this idea, it is then essential to realize that (i) contextual inference is a non-classical phenomenon, and (ii) it agrees with relativistic causality, as we explain now. 
\vskip 1mm

(i)  in classical physics, condition (PC)  as defined above is verified, and Bell's factorization condition follows. But in quantum physics, knowing  Alice's measurement  and  result allows her to predict more, without invoking any action at a distance. This is because  $\lambda\equiv \psi$ does not tell which measurements will be actually carried out by Alice and Bob - in other words,  $\lambda\equiv \psi$ is predictively incomplete \cite{JJ,completing}. Adding this information where and when it is {\bf locally} available improves Alice's prediction about Bob's result, and Bob's about Alice's, in agreement with eq.(\ref{eq4}), showing the 
suitability of the  concept of contextual inference.  This effect does not show up in classical physics, because a classical $\lambda$ is complete; but it does show up in QM, because a quantum $\psi$ 
is (predictively) incomplete,  as long as a measurement context has not been specified (for more details see \cite{completing} and the last sections  below). 
\vskip 1mm

(ii) since contextual inference only applies to the probabilities appearing in eq.(\ref{eq4}), it does not involve any physical interaction outside light cones; therefore it obeys relativistic causality.  A typical wrong line of thinking would be to say : if Alice can predict with certainty some results by Bob (perfect correlations, obtained when $a = b$), then either Bob's result is predetermined, or there are instantaneous actions at a distance. But  this dilemma only applies in a classical framework, where particles' properties are defined in an absolute way, and Bell's inequalities do apply. In a quantum framework, Alice can make local inferences  by using additional information that is available to her, e.g. $(\lambda x a)$ in the above example; and these predictions can only be checked by accessing Bob's results afterwards, in a local and ordinary causal way. 
\vskip 2 mm

\noindent{ \it Discussion.--}
It is also interesting to draw a standard light-cone picture (see below), in order to show explicitly 
how  contextual inference may be used when the relevant information is locally available \cite{ft5}.  More precisely, this diagram allows us to separate on the one hand the localized events in space time (first the production of $\lambda$, $x$ and $y$, and then the separated read-out of $a$ by Alice, and $b$ by Bob), and on the other hand the corresponding predictions, that are inferences, not influences, so no ``action at a distance" is involved.

\begin{figure}[h]
\begin{center}	
\includegraphics[width = 7.5cm]{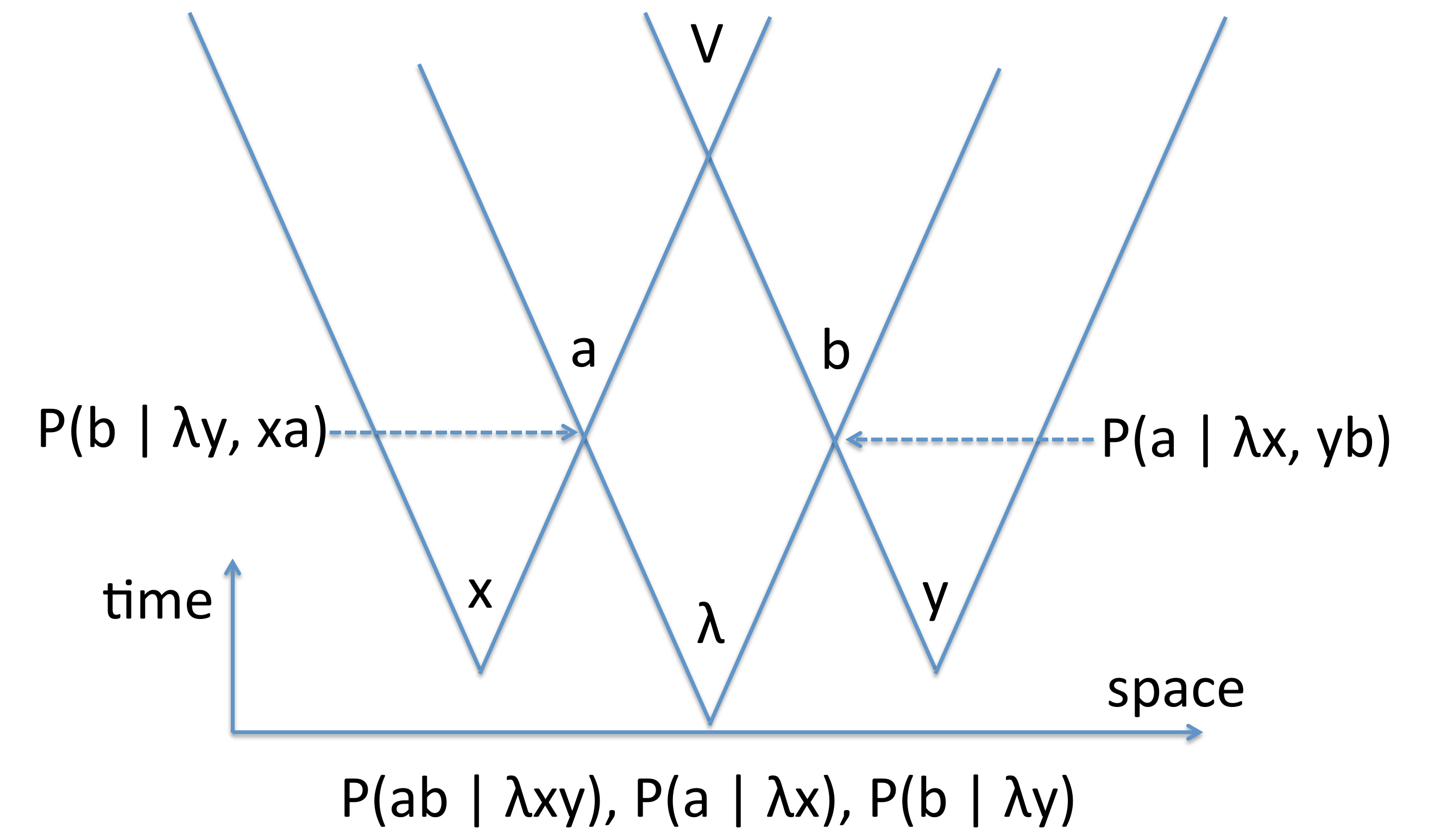}
\caption{Light-cone picture of the EPR-Bohm-Bell scheme. The photon pair is generated at the bottom of the middle cone, and is described by $\lambda$.  The measurement settings $x$ and $y$ are chosen by Alice and Bob in separated light cones. The earliest time for generating the results $a|x$ and $b|y$ are at the intersections of the light cones, and this is also when Alice's probability $P(b|y\lambda,xa)$  about Bob's result, and Bob's probability $P(a|x\lambda,yb)$ about Alice's result become available (dashed arrows). These probabilities result from a contextual inference, which respects relativistic causality and does not entail any action or influence between Alice and Bob. The resulting predictions can be effectively checked in the verification zone V, in the common future of all light cones.}
\end{center}
\label{figw}
\end{figure}

Another remark may be useful: as suggested by the light cones pictures above, one may consider that $x$ and $y$ are also issued from independent random processes in variable spaces $X$ and $Y$, as it is done in loophole-free Bell tests \cite{Aspect}. Then the global probability writes
\beq
P(x a , y b) = \sum_{\lambda \in \Lambda, x \in X, y \in Y} P(ab|xy\lambda) P(\lambda) P(x) P(y) \label{eqs1}
\eeq
where $P(ab|xy\lambda)$ is given by eq.(\ref{eq4}) as before. 
Taking $\Lambda = \{\lambda \}$, $X=\{x_1, x_2\}$ with $P(x_1)=P(x_2)=1/2$, 
$Y=\{y_1, y_2\}$ with $P(y_1)=P(y_2)=1/2$  as in a usual Bell test, 
one gets  $P(x a , y b) =  P(ab|xy\lambda)/4$.  

Correspondingly, the random variable $(x a , y b)$ may take 16 mutually exclusive values, not 4, and Bell's inequalities cannot be written anymore. 
Actually, Bell's reasoning requires to calculate  the correlation functions $E(x,y) = \langle a b \rangle_{x,y}  $ by using $P(ab | xy)$, not $P(x a , y b)$, so that the four different measurements apply to the same sample space $\Lambda$. This means implicitly that  $\lambda$ completely carries the pairs' properties (and the measurement results can be predicted from the knowledge of $\lambda$ alone), as it would be the case in classical physics. But this is counterfactual \cite{Laloe,Peres} with respect to the quantum approach, where $\{\lambda,x_1,y_1\}$, $\{\lambda,x_1,y_2\}$, $\{\lambda,x_2,y_1\}$, $\{\lambda,x_2,y_2\}$ are four different situations 
which should not be merged within a single $S$ value, contrary to Bell's reasoning \cite{Bell,BCH}. This is another way to tell that $\psi$ is not complete, and requires a context specification to be turned into an actual probability distribution. 
\vskip 3mm

\noindent {\it Completing $\psi$ ? --} A second major new step is to answer the question: If  $\psi$ is not complete, does it tell anything concrete by itself ? It does, because it indicates a set of contexts, corresponding to all the observables including $\psi$ as an eigenvector,  where the associated measurement result (eigenvalue) is predictable with certainty. In recent papers \cite{csm1,csmBell,csm3,csm4a,csm4b} we have introduced a framework which makes a careful distinction between the usual $\psi$ without a context, and the physical state within a context, called a modality (see also Appendix). In this langage $\psi$  is associated with an equivalence class of modalities, called an extravalence class \cite{csm3}: whereas the modalities are complete, because they are properties of a system within a context, $\psi$ is not, because the context is missing by construction.  This gives a nice outcome to the Einstein-Bohr debate, by confirming the incompleteness of $\psi$ \cite{EPR}, and by telling also how to complete it: one should  add the context - that actually fits with the ``very conditions" required by Bohr's answer  \cite{Bohr,Laloe,debate}. 
\vskip 1 mm

\noindent {\it Conclusion.--}
Summarizing, the violation of Bell's inequalities by quantum theory and experiments \cite{Aspect}  can be explained if one takes into consideration contextual inferences, and these in turn are ultimately allowed by the predictive incompleteness of the quantum state : getting actual probabilities for measurement results requires to specify a measurement context. Contextual inferences correspond to what is usually called ``quantum non locality'', but they are not related to locality in a relativistic sense, but rather to {\bf the specifically quantum condition that requires to attribute physical properties to systems within contexts.} The implications on the (in)completeness of QM are discussed in more details in \cite{completing}, but a few comments are in order: 
\vskip 1mm

\begin{itemize}
\item In the above we argue that $\psi$ is predictively incomplete, but not that QM is incomplete in the sense of being erroneous. There are many practical ways to complete it, by reintroducing the context either ``by hand" (like in usual textbook QM), or in a more formal way by using algebraic methods \cite{completing}. 
\vskip 1mm
\item The predictive incompleteness of $\psi$ is general, and not limited to entangled states. This is because the measurement context is required to get actual probabilities, or said otherwise, that one cannot define a full consistent set of {\bf classical}  probabilities applicable to any result in any context \cite{ft6}. This is only possible for each given context, as an immediate corollary of Gleason's theorem \cite{csm4b}. 
\vskip 1mm
\item
In this article we enforced  (EL) at the beginning, and explained how (PC) can be violated by a non-deterministic theory, without any conflict with RC. On the other hand, deterministic theories must agree with (PC), and therefore have to violate (EL) to be compatible with the observed violation of Bell's inequalities; an example of such a theory is Bohmian mechanics.  Generally speaking, if (EL) is rejected more care must be taken in order to avoid an explicit violation of special relativity \cite{norsen}.
\vskip 1mm
\item 
Here we considered the standard version of Bell's theorem, but many other inequalities may be obtained in the general framework of ``local realism". It would be interesting to look whether the violation of such inequalities is generally due to a violation of (PC); this may be the topic for further work (see Appendix for three-particle entanglement). 
\vskip 1mm
\end{itemize}

Finally, it is interesting to note that in \cite{ETJ}  Jaynes did not spell out neither the ``nonlocal" nor the ``incompleteness" option,  though he did all the calculations above. In our opinion, this is because he could not give up the classical idea that particles should be described independently of their contexts. In order to admit the idea of  contextual inference an intellectual quantum jump is required, to accept that in quantum mechanics one has to take into consideration both the systems {\bf and}  the contexts in which they evolve. A simple way not to forget this requirement is to postulate that  the ``object'' carrying well defined properties is a composite : a (quantum) system {\it within} a (classical) context \cite{completing,CO2002,csm1,csmBell,csm3,csm4a,csm4b,debate,myst}. 
\vskip 2mm

{\bf Acknowledgements.} 
The author deeply thanks Franck  Lalo\"e, Roger Balian and Nayla Farouki for many useful discussions, not meaning that they endorse whatever is written above. 




\pagebreak 

\begin{appendix}

\section*{Appendix}

In this section  we address two independant topics :

1. Summary of the main arguments, in either the standard QM or the CSM point of view.

2. Discussion of three-particle entanglement, with respect to Predictive Incompleteness. 

Some references from the main text are cited again, so this section can be read independently. 

\vskip 3mm

\subsection{1. Summary of the main arguments.}   
\noindent {\it General arguments.--} In the text above we presented a general probabilistic framework  inspired from \cite{ETJa,JJa,BJa} and we saw that, assuming the causal independence of the pair emission and measurements, the hypothesis of Bell's theorem can be split in two different sub-hypotheses : 
\vskip 2 mm

\noindent {\bf Elementary Locality (EL) :} At the most elementary level ``$\lambda"$ allowed by the theory under consideration, the probability distribution of the result $a$ of a measurement  $x$ by Alice cannot depend on the choice of a measurement $y$ done by Bob in a remote place (same thing with Bob's result $b$) : $P(a|xy\lambda) = P(a|x\lambda)$ and $P(b|xy\lambda) = P(b|y\lambda)$.  
\vskip 2 mm

\noindent {\bf Predictive Completeness (PC) :}   Again at the ``$\lambda"$ level, Alice knowing the result $a$ of her measurement $x$ cannot help her to get a better inference about Bob's result $b$ (and same thing by exchanging Alice and Bob) :  $P(a|xy\lambda b) = P(a|xy\lambda)$ and $P(b|xy\lambda a) = P(b|xy\lambda)$. 
\vskip 2 mm

Taken together  the conditions (EL)  and (PC) entail the factorization $P(ab|xy\lambda) = P(a|x\lambda) P(b|y\lambda)$, and therefore lead to Bell's inequalities \cite{Bella,BCHa}.  As a consequence, a theoretical framework violating Bell's inequalities must reject at least one of the two hypotheses.  

For instance, quantum mechanics (QM) agrees with (EL), which can be seen as a consequence of relativistic causality, but rejects (PC). This means that the description given by $\lambda$ (or $\psi$ in the quantum case) is {\bf not} complete, and that knowing $(xa)$ does bring something new to Alice : then condition (PC) can be violated, and Alice  can make  a  ``contextual inference'' about Bob's result. 

The predictive incompleteness of $\psi$ is actually not a big surprise, and it appears implicitly or explicitly in many presentations of QM, either based on the standard formalism \cite{baliana}, or using other approaches \cite{mysti}. The respective lines of arguments are as follows :
\vskip 3 mm

\noindent {\it Arguments  based on the QM  formalism \cite{baliana}.}
\vskip 1 mm

\begin{itemize}[label=\textbullet, leftmargin=*, itemsep=1pt]
\item \noindent QM is a fundamentally probabilistic theory: this is a consequence of the non-commutation of observables; 
\vskip 2 mm
\item The ``quantum state" (pure state $\psi$ or mixture $\rho$) is predictively incomplete, because by itself it does not provide a normalized probability distribution over a set of mutually exclusive events; 
\item From a physical point of view, 
$\psi$  or $\rho$ can be completed by specifying a measurement context, i.e. a macroscopic apparatus, in order to define a set of mutually exclusive events given by the apparatus outcomes. 
\item Once a context is given, $\psi$  or $\rho$ provides the relevant set of probabilities; this applies in any possible context, but only one at a time (predictive  incompleteness).
\end{itemize}

\noindent {\it Arguments based on the CSM formalism \cite{mysti}.}
\vskip 1 mm

\begin{itemize} [label=\textbullet, leftmargin=*, itemsep=1pt]
\item QM is a fundamentally probabilistic theory : this is a consequence of contextual quantization  \cite{csm1a,csm4ba};
\item Nevertheless, QM allows measurements results to be predicted with certainty, either by repeating them in the  same measurement context (this defines a modality), or by observing fully connected results betwen different contexts (this defines an equivalence classe of modalities, called an extravalence class);
\item Associating $\psi$ (a mathematical object) to an extravalence class yields Born's law from Gleason's theorem;
\item By construction $\psi$ is predictively incomplete  because it is associated with an extravalence class and not with a modality, so the context is missing. 
\end{itemize}

Then the violation of Bell's inequalities and all similar effects (resulting from ``local realism") appear as a consequence of the predictive incompleteness  of $\psi$, and have no conflict whatsoever with relativistic causality. There is no influence at a distance, but only inference at a distance, within a non-classical framework that is  fundamentally probabilistic  (this is the starting point of both lines of arguments above). 

In the CSM point of view, Quantum Mechanics relies on a non-classical ontology, where physical properties are attributed to physical objects consisting of a system within a context, i.e. an idealized measurement apparatus. Such physical properties are called modalities, and a modality belongs to a specified system within a specified context, which is described classically. Loosely speaking, the mathematical description of a modality includes both a usual state vector $| \psi \rangle$, and a complete set of commuting operators admitting this vector as an eigenstate. Though it may appear heavier at first sight, this point of view eliminates a lot of troubles about QM, and (in some sense !) it can be seen as a reconciliation between Bohr and Einstein in their famous 1935 debate \cite{debatea}. 

\subsection{2. Discussion of three-particle entanglement.}
In the main text  the argument has been based on Bell's theorem, but it is also interesting to consider three-particle entangled states, where the conflict between QM and local realism is even more straightforward, as we will see now. 
The basic argument has been introduced by Greenberger, Horne and Zeilinger \cite{GHZ} and it has been developed by many authors, both theoretically and experimentally \cite{Pan,Erven}; here we will follow mostly Mermin's presentation in \cite{Mermin1,Mermin2}.  

So let us consider measurements carried out on three spin 1/2 particles, each one with a spin operator $\vec S = \frac{\hbar}{2} \vec \sigma$, where the three Pauli matrices $( \sigma_x,  \sigma_y,  \sigma_z)$ are denoted as  $( X,Y,Z)$ for simplicity.  The $2 \times 2$ unitary matrix is denoted as $I$, and ordered products such as $XYZ$ mean the measurements of $ \sigma_x,  \sigma_y,  \sigma_z$ on the first, second and third particle, respectively given to Alice, Bob and Charlie. 

A basic feature is that the operators $XYY$, $YXY$, $YYX$ all commute together, and have 8 distinct eigenvalues $(\pm 1,\pm 1,\pm 1)$ (check it !). Therefore they define a complete set of commuting operators (CSCO), where the eight orthogonal eigenvectors are defined by the eight eigenvalues. Another feature is that $XXX$  also commute with the 3 previous operators, and one has the algebraic relation $XXX = - (XYY).(YXY).(YYX)$.  
Correspondingly, the eigenvalue of $XXX$ is $(-1)$ for the $(+1,+1,+1)$ eigenvector in the previous basis, that we will take as the initial state and denote as $\psi_{in}$ \cite{ft}.

On the other hand, denoting $x = \pm 1$ (resp. $y = \pm 1$)  the result of measuring  $X$  (resp. $Y$), 
on each spin, one has $xyy=yxy=yyx=+1$ 
in state $\psi_{in}$, and since $y^2 = 1$  the product  $xxx$ is $+1$ also.  But assuming (EL) and (PC) the 
value of each $x$ should not depend on the other measurements being $X$ or $Y$, 
and one gets a full contradiction for $\psi_{in}$, between measuring $XXX$ (and thus getting -1) 
and deducing $xxx$ from the separate measurements of $XYY$, $YXY$ and $YYX$ (and thus getting +1).
%
\begin{figure}[h]
\begin{center}	
\includegraphics[width = 9.cm]{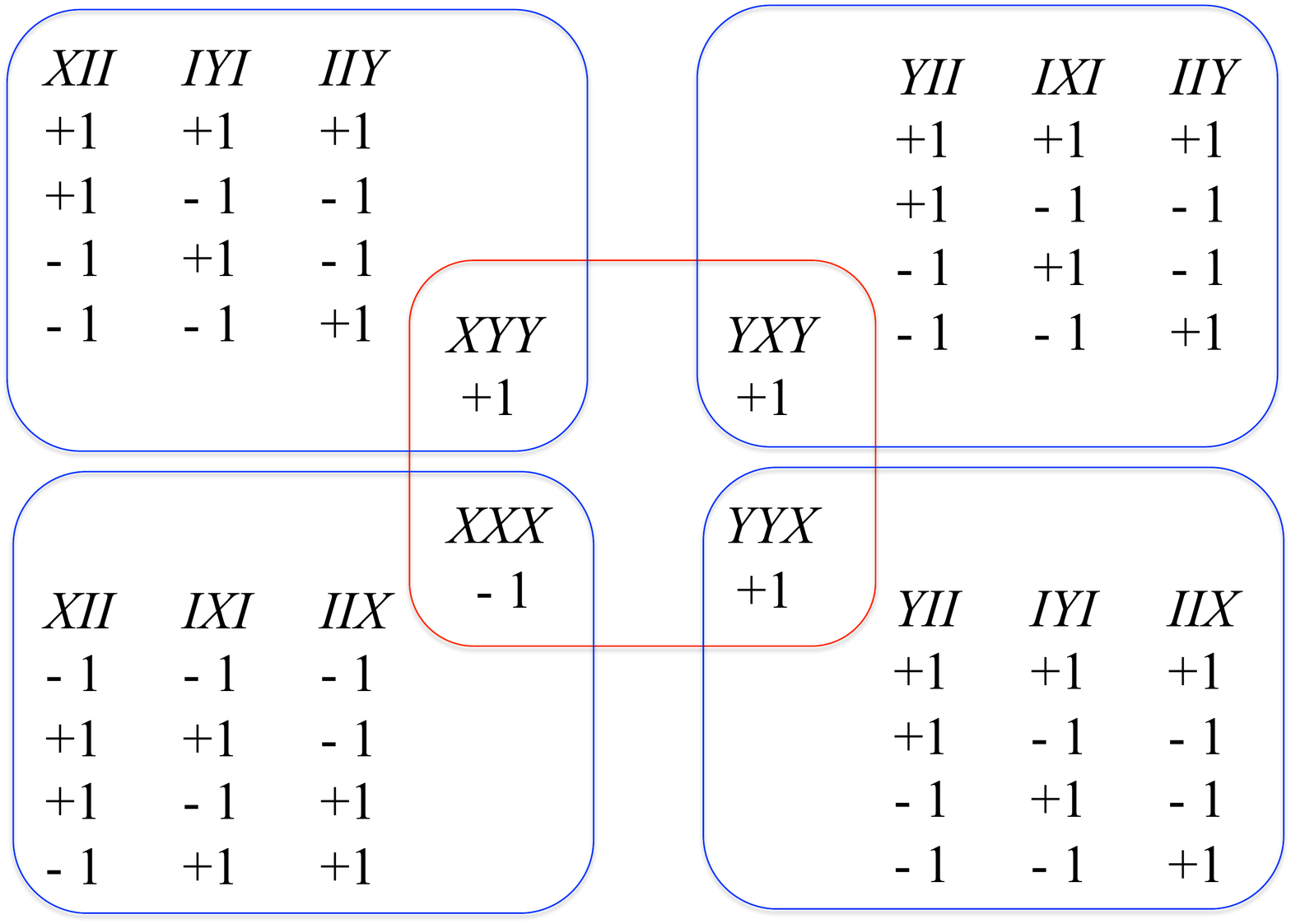}
\caption{Illustration of the complete set of commuting observables considered here. The operators $X,Y,Z$ correspond to the Pauli matrices $\sigma_x, \sigma_y, \sigma_z$ and $I$ to the identity.  Three operators in a set are enough, but a fourth commuting one is added in each group.  The possible results are indicated for the initial state $\psi_{in}$ (GHZ state), which is an eigenstate of the CSCO at the center. Note that the missing operators  $XXY$, $XYX$, $YYX$, $YYY$ all give random results in the state  $\psi_{in}$. }
\end{center}
\label{figg}
\end{figure}

For a better understanding, Fig. 2 displays the relevant CSCO for the considered situation. The previous operators are at the center, whereas $XYY$ for instance is commuting with the CSCO $(XII, IYI, IIY)$, which is incompatible with $(YII, IXI,IIY)$ and  $(YII,IYI,IIX)$. In this situation (PC) requires that given the previous initial joint  eigenvector $\psi_{in} \equiv (+1,+1,+1)$ no further measurements by Alice, Bob or Charlie can improve the inferences they can make about the results to be obtained by others. But this is clearly not the case in QM, since for instance  Alice and Bob can predict Charlie's result with certainty, by sharing their measurement results.  
\\

This issue can be further illustrated by deriving inequalities in a local realistic framework, as it was done in  \cite{Mermin2}. 
The hypothesis needed from local realism is 
\beq P(abc | uvw \lambda) = P(a | u \lambda)P(b | v \lambda)P(c | w \lambda) \label{LR} \eeq
whereas  the rules of inference together with elementary locality (EL) for 3 particles provide
\bea
P(abc|uvw\lambda) 
&=& P(a|u \lambda )P(b|uv\lambda a)P(c|uvw\lambda a b) \nonumber \\
&=& P(a|u\lambda )P(b|uvw\lambda a c) P(c|uw\lambda a)  \nonumber \\
&=& P(a|uvw\lambda bc ) P(b|v\lambda)P(c|vw\lambda b)  \nonumber \\
&=& P(a|uv\lambda b ) P(b|v\lambda)P(c|uvw\lambda ab)  \nonumber \\
&=& P(a|uvw\lambda bc )P(b|vw\lambda c) P(c|w\lambda)   \nonumber \\
&=& P(a|uw\lambda c )P(b|uvw\lambda ac) P(c|w\lambda)   \nonumber
\eea
These rules are fulfilled by QM, and each pair of equations applies respectively to Alice, Bob and Charlie.  More precisely, with $a,b,c = \pm1$ and $u,v,w = 0$ for $X$ and $\pi/2$ for $Y$, one has 
$$P(abc|uvw\lambda)  = (1 + abc \cos(u+v+w))/8$$ and all other probabilities are $1/2$ except 
$P(a|uvw\lambda bc )  = P(b|uvw\lambda a c)  = P(c|uvw\lambda a b) = (1 + abc \cos(u+v+w))/2$.  This means that if $u+v+w = 0$ or $\pi$, given their results two partners can predict with certainty the result of the third one, as it can be expected from the definition of $\psi_{in}$. 
\\

On  the other hand, in order to get eq. (\ref{LR}) one needs to add Predictive Completeness (PC), which reads by assuming (EL) is fulfilled: 
\bea
P(a|u \lambda ) = P(a|uv\lambda b ) = P(a|uw\lambda c )  = P(a|uvw\lambda bc ) \nonumber \\
P(b|v\lambda) = P(b|uv\lambda a) = P(b|vw\lambda c) = P(b|uvw\lambda a c)  \nonumber \\
P(c|w\lambda) = P(c|uw\lambda a) = P(c|vw\lambda b) =  P(c|uvw\lambda ab)  \nonumber 
\eea
For instance the first line refers to predicting Alice's result, from the initial state, then by Bob, then by Charlie, and then jointly by Bob and Charlie. According to (PC) all these predictions should be the same, whereas this is clearly not the case for QM, because $\lambda \equiv \psi_{in}$ is predictively incomplete : a new measurement in a new context provides a better contextual inference. Again, this does not imply any influence at a distance, because (EL) is always fulfilled, but only an inference at a distance, in a non-classical probabilistic framework.

\end{appendix}

\begin{thebibliography}{36}
\bibitem{options} At various points we will enforce some (reasonable) hypothesis, whereas the opposite option might be 
  taken. Such deliberate choices will be indicated in footnotes.
\bibitem{ft1}
It must be pointed out that the conditioning events $(\lambda x a )$ and $(\lambda y b )$ are meaningful both for QM and RC, as it will be discussed in more detail below.
\bibitem{ft3}
This ``elementary locality"  condition is not accepted by Bohmian mechanics (BM), which agrees with relativistic causality only after averaging on $\lambda$. Thus our initial requirement of fulfilling RC at every step - related to the universality of the RC rule - is not compatible with  BM; see also the Conclusion section.
\bibitem{ft5}
Our use of relativistic light cones should be taken with a grain of salt:  although the scheme on the figure does show causal connections 
valid  when $c$ is an invariant, it supports probability distributions and random events, e.g. measurements yielding $a$ and $b$, whereas the standard Minkowski  description involves deterministic events.
\end{thebibliography}
\begin{thebibliography}{21}
\bibitem{ft} The state $\psi_{in}$ is the usual GHZ state defined as $( | +++ \rangle - | - - - \rangle)/\sqrt{2}$, where $|\pm \rangle$ are the eigenstates of $\sigma_z$  \cite{GHZ} .
\end{thebibliography}
\end{document}